\documentstyle[pra,aps]{revtex}
\begin{document}
\draft

\title{PROPER  TIME  DERIVATIVES  IN QUANTUM MECHANICS\footnote{Published  in
Physical Review A {\bf 51}, 96-103 (1995)}} 

\author{Juan  P.  APARICIO, Fabi\'an H.   GAIOLI,  and  Edgardo  T.    GARCIA
ALVAREZ\footnote{Emails:  gaiol@iafe.uba.ar, \ \ galvarez@dfuba.df.uba.ar}}  

\address{{\it Departamento  de  F\'{\i}sica,  Facultad  de Ciencias Exactas y
Naturales,   \\   Universidad  de  Buenos  Aires,  1428  Buenos  Aires,
Argentina.}}

\date{\today}

\maketitle

\begin{abstract}

Several quantum proper time derivatives are obtained from the Beck one in the
usual framework of relativistic  quantum  mechanics  (spin $1/2$  case).  The
``scalar Hamiltonians'' of these derivatives  should  be  thought  of  as the
conjugate variables of the proper time.    Then,  the Hamiltonians would play
the  role  of  mass  operators, suggesting the  formulation  of  an  adequate
extended  indefinite  mass  framework.  We propose and  briefly  develop  the
framework corresponding to the Feynman parametrization of the Dirac equation.
In such a case we derive the other parametrizations known  in the literature,
linking  the  extension  of  the  different  proposals of quantum proper time
derivatives again. 

\end{abstract}

\vskip 2cm
Pacs number: 03.65.Pm

\narrowtext
\twocolumn

\section{Introduction}

The proper  time formulations of relativistic quantum mechanics (RQM) present
some advantages with  respect  to  the  usual  theory.  The introduction of a
``quantum proper time''\footnote{This name  is also used in the literature to
refer to any invariant evolution parameter which under certain conditions can
be related to the classical proper time.} reestablishes in quantum theory the
symmetry between space and time required by  the special theory of relativity
by elevating the coordinate $x^0$ to the rank of operator.  This also permits
us  to recover a fundamental concept (a Lorentz scalar  evolution  parameter,
the so-called ``proper time'') lost in the standard RQM.   Thus  putting in a
parallel  way  both  RQM  and  non-RQM,  it allows us to use  the  well-known
properties  of  this last theory.  The main idea of a proper  time  formalism
consists in  taking  states  that evolve with a Schr\"odinger equation, whose
``scalar Hamiltonian'' plays  the  role  of  a  mass  operator.  The standard
theory can be recovered  for  definite  mass  states.    Thus  this framework
provides  a  natural  theoretical  basis    for    a  mass  operator  concept
\cite{enatsu}.    Furthermore,  it allows solving  the  localization  problem
\cite{kalnay} by means of an extension of the Poincar\'e algebra, including a
four  ``position'' operator \cite{johnson}.  The St\"{u}ckelberg \cite{stuck}
interpretation  of    antiparticles    naturally   arises  in  the  formalism
\cite{letter}, which enables  us  to circumvent other related difficulties of
the standard RQM \cite{thaller},  without  appealing  to quantum field theory
\cite{davidon,teitel,barut0,steph}.

The idea of elevating the  time  coordinate $x^0$ to the rank of operator and
of introducing the quantum analog of  the  classical  proper time in RQM goes
back to Dirac's earlier works \cite{dir26}.  These works have been forgotten,
probably because Dirac himself did not insist on this point in his celebrated
paper  of  1928  \cite{dir28}.    However,  from the pioneer  works  of  Fock
\cite{fock}   and  St\"uckelberg  \cite{stuck},  many  authors  have  made  a
considerable  effort   to  develop  a  quantum  theory  with  a  proper  time
\cite{beck,nambu,feynman,schwin,davidon,enatsu,szamosi,peres,corben,fgood,cooke,moses,johnson,bk,chang,roman,rd,rd12,fsf,fsf12,hostler,hete,barut,kypri,droz,afa,sonego}.
In spite of  this,  the problem is still open because, among other questions,
there  exist  different proposals  and  usually  the  authors  present  their
versions omitting the relation to  other  approaches.   The interpretation of
the  additional  parameter  also  appears  as    a    conceptual   difficulty
\cite{letter}.  These topics are the core of this work.

In the majority of the contributions the  quantum proper time is a $c$ number
scalar   parameter  \cite{mac}.    Many  parametrizations  or  proper    time
derivatives have been used with some satisfactory results in each case.  Some
of  them  have  used  the formalism of the standard RQM,  while  others  have
extended it in several nonequivalent degrees.    

The influence of Fock's \cite{fock} and St\"uckelberg's  \cite{stuck}  papers
is manifest;  most authors have considered wave  equations  with second-order
space-time        derivatives        (square        mass           operators)
\cite{nambu,davidon,enatsu,cooke,roman,rd,rd12,fsf,fsf12,hostler,kypri,droz,steph,sonego}.
In  this  line,  we  can  also  mention  the  classical  paper  of  Schwinger
\cite{schwin},  about  the  vacuum polarization and  the  electron  anomalous
magnetic moment.  However, in this celebrated  paper, the author makes use of
the  St\"uckelberg  proper  time  formalism  without  stressing the  physical
interpretation of it.  Later on, Roman {\it et  al.}  \cite{roman} have shown
that  such  a  parametrization  can  be  seen  as  a  representation    of  a
five-dimensional  Galilei group, introducing a {\it universal length} $l$ (in
fact, this  parametrization is essentially the Schwinger one if the evolution
parameter is rescaled by choosing $l=2$).

On the other hand, Feynman \cite{feynman} introduced a fifth parameter in the
argument of the spinorial  wave  functions, considering the first-order Dirac
equation  \cite{beck,feynman,szamosi,peres,bk,chang,barut,afa}.  His approach
to QED by means of the parametric Green's functions has many points in common
with Schwinger's approach \cite{s2}.

At this point  it  is  natural  to  ask  which  is the relation among several
results.  In this  work  we investigate such a relation for the spin 1/2 case
and  show  why a one-particle  indefinite  mass  theory  is  the  appropriate
framework.    This  is  the  case   of  Feynman's  \cite{feynman},  Johnson's
\cite{johnson}, and Schwinger's \cite{schwin} proposals.  With  the  help  of
physical restrictions we demonstrate that the first two  parametrizations are
equivalent and the evolution parameter introduced through them is  reduced to
the proper time.  In contrast, the third parametrization, although  it  is  a
useful  tool,  does  not  lead to a classical theory in which  the  evolution
parameter  is  reduced  to  the proper time.  Thus the indefiniteness in  the
choice  of  a formalism whose evolution parameter was a suitable extension of
the classical proper time is removed.

In Sec.    II  we  show  the relation among different proper time derivatives
proposed in the  literature  in  the  framework of the standard definite mass
theory.  We propose,  in  Sec.  III, an indefinite mass formalism and analyze
the corresponding relations in such  a  context.  For the sake of simplicity,
we consider only, the minimal coupling case for electromagnetic interactions.
The  extension  to more general interactions will  be  briefly  discussed  in
Appendix B.

\section{Proper time in the standard Dirac theory}

Although the introduction of  a  quantum  proper  time  naturally leads to an
indefinite  mass theory in a  direct  way  (see  Sec.    III),  the  standard
framework  of  RQM (a definite mass  theory)  was  frequently  used  by  many
authors.   In this framework we can  only  work  in  a  ``formally  covariant
manner'' since the temporal coordinate $t$ is a  $c$  number.  In such a case
the  inclusion  of  the  proper time\footnote{The analogy with the  classical
counterparts  derived  from  the  equations  of motion is the origin  of  the
inaccurate name given to such a parameter.} is made through a  quantum proper
time  derivative  of  the  dynamical  variables.    This  derivative  is  the
counterpart of the ordinary time  derivative  of  a dynamical variable $q$ in
the Heisenberg picture,\footnote{We use natural units  $\hbar  =c=1$  and the
notations and conventions are those of Messiah \cite{messiah}.} which reads 

\begin{equation}    
\frac{dq}{dt}=       i[H,q]+\frac{\partial    q}{\partial
t},\label{dto}
\end{equation} 
with  $H$  the  Dirac  Hamiltonian,  i.e., for minimal  coupling,  $H  ={\vec
\alpha}\cdot {\vec \pi} + \beta m_0 + e\phi $.    If  $H$  were replaced by a
``scalar operator,'' we would obtain a quantum proper time derivative.    For
example,  the Beck derivative \cite{beck} for a variable $q$, which does  not
explicitly depend on the proper time $s$, is defined by

\begin{equation}    
(\frac{dq}{ds})_B\equiv    -i[{\cal      H},q],\label{db}
\end{equation} 
where,  for  minimal coupling, ${\cal H} = \gamma^\mu\pi_\mu$ (with $\pi _\mu
=p_\mu -eA_\mu$  and $ p_\mu =i\partial _\mu $).  This is the natural choice.
However, in most works the Fock \cite{fock,nambu} derivative is used:

\begin{equation}    
(\frac{dq}{ds})_F\equiv-\frac{i}{2m_0}[{\cal H}^2,q]\label{df}.
\end{equation} 
This  is  due  to  the  following.    While  equations of motion with  formal
classical  analogy  are  obtained  by  means of the Beck as well as the  Fock
derivatives,  Beck's  results  do  not  seem to have such an analogy at first
glance.   This  happens  because  the  equations  of motion actually describe
``particle-antiparticle states''(classified by  the  sign of the mass) having
covariant  {\it  Zitterbewegung}  \cite{zit}.\footnote{This  fact  cannot  be
interpreted in the context developed in this section.  The adequate framework
is an indefinite mass theory that  we  will introduce in Sec.  III.} Thus the
nonpreference  for  the  Beck  derivative  is  due   to  the  fact  that  the
corresponding  classical  theory of spinning particles \cite{bhabha} is  less
known.  We leave the discussion of this point  for a separate paper, where we
will study these equations in more detail \cite{spin}.

In addition to the works of Fock and Beck, in the 1960s, Corben \cite{corben}
proposed

\begin{equation}
(\frac{dq}{ds})_C\equiv \beta \frac{dq}{dt}\label{dc}
\end{equation}
as  the  quantum  proper  time derivative, inspired in the classical relation
between $ds$ and  $dt$,  by identifying $[(1-v^2)^{-1/2}]_{op}$ with $\beta$.
However, the Corben derivative does not satisfy Leibniz's rule \cite{bk}.  He
obtained some plausible results, but his hope to obtain a parallelism between
classical  and quantum equations of motion  of  spinning  particles  was  not
completely successful.  

In 1961, Fradkin and Good \cite{fgood} considered

\begin{equation}
(\frac{dq}{ds})_{FG}\equiv \langle \beta\rangle
\frac{dq}{dt}\label{dfg}
\end{equation}
as a quantum proper time derivative, where $\langle \beta\rangle $ stands for
the mean value  of  $\beta$.   This proposal satisfies Leibniz's rule and has
desirable properties in the classical limit.  But it is strongly dependent on
the quantum state \cite{bk}.

Let us also remark that,  up  to  now,  the  results  obtained  from the last
derivatives are, in general, different (see Appendix A).  In the following we
will show a connection among them.   So,  even  if some derivatives mentioned
above were not good candidates in general, their results can be obtained from
the Beck derivative.  

Let  us start by considering the Beck  and  the  Corben  derivatives.    From
definition (\ref{db}) it follows that a dynamical variable $q$ satisfies

\begin{equation}
(\frac{dq}{ds})_B=\beta \frac{dq}{dt}+i[q,\beta ](i\frac{\partial
}{\partial t}-H),
\label{prbc}
\end{equation}
where we have used that ${\cal H}$  can  be rewritten as\footnote{Notice that
this  identity  is valid not only for minimal  coupling,  but  also  for  the
general case discussed in Appendix B.} $\beta(i\partial _t -H) + m_0$ and Eq.
(\ref{dto}).

If $[\beta ,q]=0$, the Corben and the Beck derivatives  are  equivalent.   In
general $[\beta ,q]\neq 0 $, but these derivatives coincide when  are applied
to solutions $\phi(x)$ of the Dirac equation:  

\begin{equation}
(\frac{dq}{ds})_B\phi (x )=\beta \frac{dq}{dt} \phi (x).\label{rbc}
\end{equation}
Let us note that, although the Corben derivative is a nonacceptable proposal,
it works  under  some  conditions.    For  example,  the relation (\ref{rbc})
explains why, in  some  cases,  Corben  had  to  restrict  his  equations  to
solutions of the Dirac equation.

With the help of (\ref{rbc}) we can prove  that,  in the semiclassical limit,
the  Beck  and  the  Fradkin-Good  derivatives  are  equal  in  mean  values.
Moreover, the Fradkin-Good derivative was introduced precisely in this limit.
In fact,  considering  the  standard  semiclassical  coherent  states,  which
factorize the mean  value  of  the  product  of  operators  to first order in
$\hbar$ \cite{yaffe}, we have

\begin{equation}
\langle (\frac{dq}{ds})_B\rangle =\langle (\frac{dq}{ds})_C\rangle=\langle
\beta\rangle \langle  \frac{dq}{dt}\rangle .\label{rbfg}
\end{equation}

Before establishing the connection between the Fock and the Beck derivatives,
let us make a little  digression  about  a  new  set of variables.  They were
introduced  by  Bunge  and K\'alnay \cite{bk}  in  an  attempt  to  obtain  a
relativistic generalization of the Ehrenfest theorems.   We will call them BK
variables, which are defined by

\begin{equation}
Q\equiv q+\frac{i}{2m_0}(\frac{dq}{ds})_B,\label{bv}
\end{equation}
where $q$ is a usual Dirac variable and  $m_0$  is  the mass of the particle.
These variables were sometimes discussed in the literature.  Thus, e.g.,

$$X^\mu =x^\mu +\frac{i}{2m_0}\gamma ^\mu \nonumber$$ is a formally covariant
position \cite{qed,bunge,szamosi,bk} whose Beck derivative  

\begin{equation}
(\frac{dX^{\mu}}{ds})_B=  \frac{\gamma^{\mu}}{m_0}  (m_0-\gamma^{\nu}
\pi_{\nu}) +\frac{\pi^{\mu}}{m_0},  \label{XB}
\end{equation}
is a BK variable and has curious properties  \cite{giamb}.    The BK variable
corresponding  to the spin $\sigma^{\mu\nu} = (i/2) [\gamma^\mu, \gamma^\nu]$
is the well-known Hilgevoord-Wouthuysen \cite{hw,kolsrud} spin tensor,

$$\Sigma ^{\mu \nu }=\sigma ^{\mu \nu }-\frac{i}{m_0}(\gamma ^{\mu }\pi ^{\nu
}-\gamma  ^{\nu }\pi ^{\mu }),\nonumber$$ proposed originally  to  achieve  a
conserved spin in the free case.

On the other hand, if we consider the polarization operator defined by Michel
and  Wightman  \cite{mw},  $t^\mu  =i\gamma  ^5\gamma  ^\mu$, the BK variable
associated with it is

\begin{equation}
T^\mu =i\gamma ^5(\gamma ^\mu -\frac{\pi^\mu}{m_0}).\label{tmu}
\end{equation}
This operator was introduced by Fradkin  and Good \cite{fgood} and it is such
that,  applied to solutions of the Dirac  equation,  it  coincides  with  the
polarization  operators  considered  by  Bargmann  and Wigner (BW)  \cite{bw}
($T_{BW}^{\mu}  =  \frac{1}{2} \epsilon^{\mu\nu\rho\lambda}  \sigma_{\nu\rho}
\frac{\pi_{\lambda}}{m_0}$)  and  by  Kolsrud  \cite{kolsrud}  ($T_K^{\mu}  =
\gamma_5  \sigma^{\mu\nu}  \frac{\pi_{\nu}}{m_0}$).\footnote{The  BW  and the
Kolsrud  operators  are  equal;    it  is  easily  checked  by  the  identity
$\frac{1}{2}  \epsilon^{\mu\nu\rho\lambda}  \sigma_{\rho\lambda}  =  \gamma_5
\sigma^{\mu\nu}$.  Using that  $\gamma^{\mu}  \gamma^{\nu}  =  g^{\mu\nu} - i
\sigma^{\mu\nu}$,  the  Kolsrud  operator  can   be  rewritten  in  the  form
$T_K^{\mu}  = i \gamma_5 (\gamma^{\mu} \gamma^{\nu}  \frac{\pi_{\nu}}{m_0}  -
\frac{\pi^{\mu}}{m_0})$.  When applied to solutions of  the  Dirac  equation,
the Kolsrud operator coincides with the operator of  Eq.    (\ref{tmu}).   We
have used  the conventions $\epsilon^{0123}=1$ and $\gamma^5=\gamma^0\gamma^1
\gamma^2\gamma^3$.}

Now, let us find  the  relation  between  the  Fock and the Beck derivatives.
Observe that 

\begin{equation}
(\frac{dX^{\mu}}{ds})_B    \phi(x)    =
(\frac{dx^{\mu}}{ds})_F\phi(x)    =
\frac{\pi^{\mu}}{m_0}\phi(x). \label{xBF}
\end{equation}
where $\phi(x)$ is a solution  of  the  Dirac  equation.   The last equality,
which comes from Eq.  (\ref{XB}),  shows  that  the BK velocity has classical
analogy \cite{bk}, as it happens in the  Fock case.  Thus, from this equation
it  is  tempting to find a general relation.    In  fact,  a  straightforward
calculation of the Beck derivative of a BK variable leads to

\begin{equation}
(\frac{dQ}{ds})_B=-\frac{i}{m_0}[{\cal H},q](m_0-{\cal H})
-\frac{i}{2m_0}[{\cal H}^2,q].\label{prfb}
\end{equation}
Then,  if  $\phi (x)$ is a solution of the  Dirac  equation,  we  obtain  the
general result

\begin{equation}
(\frac{dQ}{ds})_B\phi (x)=(\frac{dq}{ds})_F\phi (x).\label{rfb}
\end{equation}

The  Beck derivative of BK variables coincides with the  Fock  derivative  of
ordinary variables applied to solutions of the Dirac equation.   In  the next
section  we prove that, in the framework of an indefinite mass  theory,  this
relation is even closer.  Summing up, we have seen in this  section that from
the  Beck  derivative,  which  is  the  natural  extension  of  a proper time
derivative,  the  results  derived  from  all  derivatives  proposed  in  the
literature can be  obtained.   Moreover, we will see in the next section that
this derivative corresponds, in  an  adequate  framework,  to  the Heisenberg
picture of the Feynman parametrization of the Dirac equation.

\section{Proper time in indefinite mass theories}

As  we  have already anticipated in the Introduction, a quantum  proper  time
derivative  can be introduced by first principles only in an indefinite  mass
theory.  These theories have in common a Schr\"odinger equation in which  the
evolution  parameter  is  the  proper time and the role of the Hamiltonian is
played  by a scalar (mass) operator.  A general solution of this equation has
mass dispersion;  thus results the name of such theories.  It is important to
remark  that  most    works    have  considered  spinless  systems  and  used
parametrizations with Hamiltonians quadratic in the momenta.  
 
We shall revise the  most well-known approaches for the spin $1/2$ case.  For
example, in Schwinger's approach \cite{schwin},  the  integral representation
of  the  Green's  function  of  the   Dirac  equation  suggested  the  formal
introduction of an evolution operator in a  proper  time $\lambda$ defined by
${\cal U}(\lambda )=e^{i{\cal H}^2\lambda}$.  Thus the Heisenberg equation of
motion for a variable $q$ not explicitly $\lambda$ dependent is

\begin{equation}
\frac{dq}{d\lambda }=-i[{\cal H}^2,q ],\label{ds}
\end{equation}
which we will  call  the  Schwinger  derivative.\footnote{All  definite  mass
derivatives were distinguished with  different names, but identified with the
same  proper time parameter $s$.    However,  in  indefinite  mass  theories,
different letters are used for the  evolution  parameters  when  they are not
directly related to the proper time.} We  can  see  that,  in  the  classical
limit, the evolution parameter $\lambda$ is unrelated to the proper time $s$.
In  fact,  only  after  choosing  the  initial  conditions on  a  given  mass
shell,\footnote{Classically $\pi^{\mu}  \pi_{\mu}$  is  a constant of motion,
which  can  be    fixed    to    a    particular   value  $m_0^2$.}  we  have
$\lambda=s/(2m_0)$.  

In    recent   works  some  authors    have    returned    to    this    line
\cite{hostler,kypri,droz,sonego}  trying  to provide a theoretical  framework
for this parametrization (for the spin 0  case).    Observe that, in order to
have  a  more  direct  identification  of the evolution  parameter  with  the
classical  proper  time,  the Schwinger Hamiltonian must be rescaled  with  a
factor  $1/(2m_0)$, as in the Fock derivative (Sec.  II).    However,  as  we
shall  see  later, it is not easy to achieve such an  identification  without
violating  the indefinite mass character of the theory.  The approaches known
as relativistic dynamics (RD) \cite{rd,rd12} and four space formulation (FSF)
\cite{fsf,fsf12} are attempts  to  give  a  foundation to a formalism of this
kind.  However, similar objections arise.  

The proper time derivative  of  RD and FSF has a Hamiltonian which depends on
an {\it intrinsic mass parameter} $M$, 

\begin{equation}
\frac{dq}{d\tau}=-\frac{i}{2M}[{\cal
H}^2, q],\label{fan}
\end{equation}
not  related  {\it  a    priori}    with    the    mass    $m_0$    of    the
particle,\footnote{Strictly speaking, Eq.  (\ref{fan})  is used by RD and FSF
in the free case (${\cal H}^2  =p^\mu p_\mu$).  For the interaction case they
use  a  slightly different formalism;  see  Ref.    \cite{rd12}.}  as  it  is
desirable in an indefinite mass formalism.\footnote {Equation (\ref{fan})  is
also the case of the Roman parametrization \cite{roman} identifying  $M$ with
the inverse of the {\it universal length} $l$.} However, in  order  that  the
evolution parameter $\tau$ becomes the proper time in the classical limit  we
must  identify  (on shell)  $M$ with the mass $m_0$ {\it a posteriori}.   The
last identification  seems  to  be  an unattractive feature for an indefinite
mass theory, since  the Hamiltonian includes the information about some given
initial conditions.    

Finally,    let    us   comment    on    Johnson's    interesting    approach
\cite{johnson,chang}.  In his first  work  \cite{johnson},  he  considers all
spins  in  the  free  case  using  a  unique  Hamiltonian  ${\cal  H}_J\equiv
\sqrt{p^{\mu    }p_{\mu}}$,   which  governs  the  evolution  in    a    time
$s$.\footnote{As we shall see in Sec.  III.A,  the evolution parameter of the
corresponding  classical  theory (independently proposed by Moses for spin  0
\cite{moses}) is the proper time.} The physical states belong to the subspace
spanned  by  eigenfuctions  of  ${\cal  H}_J$  with positive eigenvalues.  In
general, we define the extension of the Johnson Hamiltonian as

\begin{equation}
{\cal H}_J\equiv \sqrt{{\cal H}^2} \equiv m_{op},\label{hj}
\end{equation}
where,  e.g.,  ${\cal H}^2=\pi^\mu \pi_\mu- (e/2) \sigma^{\mu\nu} F_{\mu\nu}$
for minimal  coupling.  The Heisenberg evolution of a dynamical variable $q$,
which  does not  explicitly  depend  on  $s$,  will  be  called  the  Johnson
derivative  

\begin{equation}
(\frac{dq}{ds })_J \equiv -i[ \sqrt{{\cal H}^2},q ]. \label{dj}
\end{equation}

We will not consider in more detail either the second-order [Eqs.  (\ref{ds})
and (\ref{fan})] or square root  [Eq.  (\ref{dj})] parametrizations.  We will
see how to relate these alternatives  to  a first-order parametrization whose
dynamics  is  governed  by  the  scalar Hamiltonian  ${\cal  H}$,  originally
proposed by Feynman \cite{feynman}.

\subsection{A  proper  time  formulation  of  RQM  based  on   a  first-order
parametrization}

The way to introduce the concept of proper time  in  the spin $1/2$ RQM would
be  achieved  within  a  framework  which  should  satisfy  at  least   these
requirements:    (a)  Dirac's  theory  must  be  somehow  included;   (b) the
equations of  motion  must  be  analogous  to a classical theory in which the
evolution parameter is  the  proper  time;    (c)  the  framework must not be
restricted  to the mass  shell.    Condition  (a)  is  Bohr's  correspondence
principle in a broad sense,  while  (b)  is the same principle applied to the
relation  between classical and quantum mechanics.    It  will  allow  us  to
identify the evolution parameter with the classical  proper  time.  Condition
(c)  is  what  we  have called an ``indefinite  mass  theory,''  which  is  a
consequence of extending the Poincar\'e algebra \cite{moses,johnson,roman} in
order to  include  a  Dirac  four-vector  position  operator  $x^{\mu}$,  the
canonical conjugate of $p_\mu$,

\begin{equation}
[x^\mu,p^\nu]=-i\hbar g^{\mu\nu}.\label{cerra} 
\end{equation}
In fact, just at the  classical  level,  we  can see that the mass constraint
($p^{\mu}    p_{\mu}=m^2$)  is  incompatible  with  the    Poisson    bracket
corresponding to Eq.  (\ref{cerra}).

Taking  condition (c)  into  account,  which  allows  realizing  the  algebra
(\ref{cerra}), let us consider  a  wave  function  $\Psi(x,s)$ belonging to a
linear space of spinorial functions defined in space-time.  The wave function
represents the state of the system  at a given value of the parameter $s$ and
its evolution is given by

\begin{equation}
-i\frac{d}{ds}\Psi (x, s)={\cal H}\Psi(x,s).\label{ef}
\end{equation}
The scalar Hamiltonian ${\cal H}$ plays the role of the standard  Hamiltonian
in non-RQM.  From (\ref{ef}), the evolution operator is

\begin{equation}
U(s)=e^{i{\cal H}s}.\label{oef}
\end{equation}
Going  to the Heisenberg picture, in which the evolution of the operators  is
governed by

\begin{equation}
\frac{dq}{ds}= -i[{\cal H},q],\label{dbe}
\end{equation}
we reobtain the Beck equation of motion (\ref{db}).

The eigenfunctions  of  ${\cal  H}$  are definite mass states $\phi _m $ that
satisfy the eigenvalue  equation  ${\cal  H}\phi  _m=m\phi  _m$.    They have
oscillatory behavior in $s$ and are solutions of a generalized Dirac equation
[condition  (a)].    In addition,  in  the  space  of  these  spinorial  wave
functions, we define an indefinite Hermitian form \cite{rumpf,fv,barton}

\begin{equation}
\langle \Phi |\Psi \rangle \equiv \int \overline{\Phi }\Psi d^4x,\label{pe}
\end{equation}
where $\overline \Phi =\Phi ^{\dagger}\gamma ^0 $ is the usual Dirac adjoint.

Observe that the spin variables $\gamma ^{\mu  }$,  the  spin  tensor $\sigma
^{\mu \nu }$, and the orbital variables $p_{\mu }$ and $x^{\mu }$, as well as
the  Hamiltonian,  become  Hermitian  in  the  ``scalar product'' (\ref{pe}).
Hence  the  evolution operator (\ref{oef}) is unitary  \cite{letter}.    From
this, the ``norm'' of the states is a constant of motion, i.e.,

\begin{equation}
\frac{d}{ds}\langle \Psi |\Psi \rangle =0.\label{cn}
\end{equation}
Thus the subspaces corresponding to the states with  positive,  negative,  or
null  norms  are invariant under the proper time evolution.    Moreover,  the
standard  interactions,  e.g.,  the  ones  considered  in this work, are  $s$
independent  and  then they cannot produce transitions among states belonging
to different subspaces.

Analogously  to   what  happened  in  Sec.    II,  the  equations  of  motion
corresponding to different  parametrizations  are,  in general, different too
(see Appendix A).   To establish a connection among them we will consider the
restriction  of  the formalism on  the  positive  mass  subspace,  i.e.,  the
subspace invariant under the action of  the  projector defined by\footnote{It
is a straightforward extension of the well known positive energy projector in
the standard case.  See, e.g., Ref.  \cite{messiah}.  The projector (\ref{p})
was  also  used  by  Johnson  and  Chang  (see  the   second  paper  in  Ref.
\cite{chang})  and  by  Enatsu  and Kawaguchi (see the second paper  in  Ref.
\cite{enatsu}).}

\begin{equation}
\Lambda \equiv \frac{1}{2} (1+\frac{{\cal H}}{m_{op}}).\label{p}
\end{equation}
We will see  that  with  this  restriction  the  Feynman  parametrization  is
equivalent to the Johnson  one.  First, let us note that in this subspace the
Hamiltonians ${\cal H}$ and ${\cal H}_J$ coincide,

\begin{equation}
\Lambda {\cal H} \Lambda =\Lambda {\cal H}_J\Lambda. \label{hp}
\end{equation}
Observe also that

\begin{equation}
[{\cal   H},\Lambda]=0=[{\cal  H}_J,\Lambda],\  \    \    \
\frac{d\Lambda }{ds}=0=(\frac{d\Lambda }{ds})_J,\label{dp}
\end{equation}
and therefore the action  of the projector is invariant under the proper time
evolution.  From the definitions (\ref{dbe}) and (\ref{dj}), for any variable
$q$ we have

\begin{equation}
\Lambda\frac{dq}{ds}\Lambda                    =-i\Lambda[\sqrt{{\cal
H}^2},q]\Lambda=\Lambda (\frac{dq}{ds})_J\Lambda.\label{rbj}
\end{equation}
Thus, in the subspace of positive mass,  the Beck and the Johnson derivatives
are equivalent.

Now,  let  us  take  the  classical limit in  order  to  visualize  a  closer
connection among the different parametrizations discussed at the beginning of
this section.  In this way, we will also be  able  to  identify the parameter
$s$ with the proper time.

Let us first recall that if  $A$  and  $B$  are  operators,  and $f(B)$ is an
operator function, it follows that 

$$[A,f(B)]=      \frac{1}{2}\frac{df}{dB}[A,B]+[A,B]\frac{1}{2}\frac{df}{dB}+
O(\hbar ^2).$$ Therefore, to first order in $\hbar$, we have 

$$\Lambda \frac{dq}{ds}\Lambda  =\Lambda\frac{1}{2}(\frac{1}{2m_{op}}[-i{\cal
H}^2,q]+[-i{\cal H}^2,q]\frac{1}{2m_{op}})\Lambda .$$  Then,  if we take mean
values with a positive  mass  quasiclassical  state\footnote{This  means that
$\Lambda  |\ \rangle =[1+ O(\hbar)]|\  \rangle$.}  and  consider  that  these
states  factorize  the  operator  product  up    to    the   first  order  in
$\hbar$,\footnote{The  quasiclassical  states  considered  in  this case  are
different from those considered in the standard Dirac  theory.    In  Dirac's
theory, since time $t$ is a $c$ number, these states are essentially the same
coherent states of non-RQM.  In indefinite mass theories the  role  played by
$t$ is very different.  Then, the quasiclassical states must be  generalized.
A straightforward generalization is shown in Ref.  \cite{letter}.} we obtain

\begin{equation}
\langle \frac{dq}{ds}\rangle =\langle \frac{1}{2m_{op}}\rangle \langle
\frac{dq}{d\lambda}\rangle                      =                \langle
\frac{M}{m_{op}}\rangle\langle\frac{dq}{d\tau}\rangle,\label{nrfb}
\end{equation} 
where, in  this  limit,  $m_{op}  =  \sqrt{\pi^\mu  \pi_\mu}$.  This equation
relates the Feynman  parametrization with the Schwinger, RD, and FSF ones, in
the classical limit.  Observe that RD and FSF can only recover the results of
the Feynman parametrization after choosing  the  initial  conditions  on  the
mass shell $\langle m_{op} \rangle = M$.     

From the first equality of Eq. (\ref{nrfb}), we obtain

\begin{equation} 
\frac{d}{ds}  \langle x^\mu\rangle\frac{d}{ds}\langle  x_\mu  \rangle
=\frac{\langle \pi^\mu \rangle \langle \pi_\mu\rangle }{\langle 
m_{op}\rangle^2}=1,\label{rtpc}
\end{equation}
since    $\langle    m_{op}\rangle  ^2  =    \langle    \pi^\mu\rangle\langle
\pi_\mu\rangle $ because of the factorization  property.  Thus $s$ is reduced
to the proper time of indefinite mass  states  that  follow  the  world  line
$\langle x^\mu\rangle (s)$.

We have seen that the formalism briefly sketched  in  this section permits us
to obtain the desired results included in all the  parametrizations,  without
appealing to {\it ad hoc} assumptions.  Moreover, the equations  of motion of
the main dynamical variables derived from Eq.  (\ref{nrfb}) read (in the free
case) 

\begin{mathletters}\label{plif}
\begin{equation}
\frac{d}{ds}\langle        x^\mu        \rangle       =   \frac{\langle 
p^\mu\rangle 
}{\langle m_{op}\rangle },\label{dx}
\end{equation}

\begin{equation}
\frac{d}{ds}\langle p^\mu \rangle =0\label{dp1},
\end{equation}

\begin{equation}
\frac{d}{ds}\langle \sigma ^{\mu \nu} \rangle =0. \label{ds1}
\end{equation}
\end{mathletters}
These  equations  on  shell  have  a classical analogy;  the four-velocity is
proportional to  the four-momentum, which results in a conserved quantity, as
is the case  of  the spin tensor.  This fulfills the outline of our proposal,
since from (\ref{rtpc}) and (\ref{plif}) we satisfy condition (b). 

In order to complete the picture of Sec.  II let us see how to generalize the
relation (\ref{xBF}) from an extension of the BK variables.  The BK variables
used in the definite mass theory  have no meaning in the indefinite mass case
since they depend on a particular mass  value.    One way of bypassing such a
difficulty  consists  in  making  the  heuristic substitution $m_0\rightarrow
m_{op}$.  So, we define a new BK variable as

\begin{equation}
Q\equiv q+\frac{i}{2}(-i[\frac{{\cal H}}{m_{op}},q]).\label{nbkv}
\end{equation}
For example, the position operator reads

$$X^{\mu}=  x^{\mu}  +  \frac{i}{2m_{op}}\gamma^{\mu}  -  \frac{i  p^{\mu}}{2
m_{op}^3}{\cal H}$$ in the free case.  Here an extra term appears because, in
general,    the   variable  $q$  does  not  commute    with    $m_{op}$    in
(\ref{nbkv}).\footnote{The  unification  of  the  different proposals for the
polarization operators shown in Sec.  II can be  easily  transposed  to  this
case.}

Using Eq.  (\ref{hp}) it is immediate  that, for a dynamical variable $q$, we
have 

$$\Lambda  q  \Lambda  =\Lambda  Q\Lambda  \nonumber$$  and  trivially,  with
(\ref{dp}) and (\ref{rbj}),

\begin{equation}
\Lambda\frac{dQ}{ds}\Lambda=\Lambda(\frac{dq}{ds})_J\Lambda.\label{fbg}
\end{equation}
Then, the Johnson derivative\footnote{Notice that the Fock derivative becomes
the  Johnson  one    by   means  of  the  same  heuristic  substitution  $m_0
\longrightarrow m_{op}$.} of the  Dirac  variables  coincides  with  the Beck
derivative of the new BK  variables,  which constitutes the generalization of
(\ref{rfb}).

\section{Concluding remarks}

We have seen in this paper that there  are different proposals to introduce a
quantum  proper  time  derivative  in  RQM,  but  they  are   not  completely
satisfactory.  Recalling the conditions imposed to a proper time  formulation
of RQM in Sec.  III.A, we can see that some  proposals,  e.g.,  RD  and  FSF,
satisfy  condition (a), but fail to give an adequate framework simultaneously
compatible with  conditions (b) and (c).  A natural extended framework, which
resembles the usual  non-RQM, that satisfies conditions (a) and (c) was given
in this paper by  means  of  the  Feynman  parametrization  (used by him as a
formal tool for obtaining the  famous  results  of  QED  in  a  heuristic way
\cite{feynman}).  It has possibly been  forgotten  because apparently it does
not verify condition (b).  However, we  have  demonstrated  in this work that
the  Feynman  parametrization  is  compatible with condition (b).    It  also
permits  us  to  derive  the  other  proposals  known in  the  literature  as
particular  cases,  establishing  a  connection  among  them  under  adequate
restrictions.  Thus we have shown a unified version of different proposals of
proper time derivative,  both  in  the  standard  as  well as in the extended
(indefinite mass) framework of RQM.

Summing  up,  a  unique  parametrization of the spin $1/2$  ``particle'' (the
natural  one), which includes the standard formalism of RQM as  a  particular
case, is enough in order to have an overall view of  the  different  previous
proposals.

\section*{Acknowledgments}

We are grateful to A.J.  K\'alnay for interesting and useful discussions.  We
would like to thank C.S.   Bernardou  for  reading  the  manuscript.   We are
supported by the Universidad de Buenos Aires. 

\newpage

\appendix
\section{}

In Sec.  II we stated that the results obtained through the several proposals
of proper time derivatives are, in general, different.    Let  us  see, as an
example, only the velocities derived from them.  These read

\begin{mathletters} \label{derix}
\begin{equation}
(\frac{dx^{\mu}}{ds})_B=\gamma^{\mu},\label{xB}
\end{equation}

\begin{equation}
(\frac{dx^{\mu}}{ds})_F=\frac{\pi^{\mu}}{m},\label{xF}
\end{equation}

\begin{equation}
(\frac{dx^{\mu}}{ds})_C=\gamma^{\mu},\label{xC}
\end{equation}

\begin{equation}
(\frac{dx^{\mu}}{ds})_{FG}=\langle\beta\rangle\beta\gamma^{\mu},\label{xFG}
\end{equation}
\end{mathletters} 
which confirms the above statement.

On the other hand, in the indefinite mass case  (Sec.   III) we have, for the
derivatives of the position operator, 

\begin{mathletters} \label{dervx2}
\begin{equation}
\frac{dx^{\mu}}{d\lambda}=2\pi^{\mu},\label{xS}
\end{equation}

\begin{equation}
\frac{dx^{\mu}}{d\tau}=\frac{\pi^{\mu}}{M},\label{xP}
\end{equation}

\begin{equation}
(\frac{dx^{\mu}}{ds})_J  =  \frac{1}{2}  \big(\pi^{\mu}  \frac{1}{m_{op}}  +
\frac{1}{m_{op}} \pi^{\mu}\big) + O(\hbar^2), \label{xJ}
\end{equation}

\begin{equation}
\frac{dx^{\mu}}{ds}=\gamma^{\mu}.\label{xFe}
\end{equation}
\end{mathletters} 
The corresponding statement is verified again.

\section{}

Let us comment on some extensions of the results of previous sections for the
case of interactions more general than minimal coupling.  Even if  the  great
majority  of the works about proper time have considered only the free  case,
and  sometimes  the  minimal  coupling  case,  the  interest  for  nonminimal
couplings and for nonelectromagnetic  interactions  has been revived recently
in connection with the quantum  equations  of  motion  of  spinning particles
\cite{heinz,afa,spin,color}.

Let us consider the scalar Hamiltonian

\begin{equation}
{\cal H}_R=\gamma ^\mu p_\mu - R,\label{hr1}
\end{equation}
where  $R$  is  an arbitrary operator which preserves the Lorentz  and  gauge
invariance  of  the  theory.    If  the  different derivatives are adequately
redefined, all  relations presented in this work remain valid.  It is easy to
check this fact,  since  the results have been obtained without making use of
the explicit form of $R$.

An example of a possible extension of the scalar Hamiltonian is

\begin{equation}
{\cal  H}_R=\sum_{n=0}^{\infty  }[\epsilon _n \gamma ^\mu \Box  ^n
A_\mu +\frac{i}{2}  \mu  _n  \gamma  ^\mu \gamma ^\nu \Box
^n(\partial  _\mu  A_\nu -\partial _\nu A_\mu)], \label{a1}
\end{equation} 
which was originally proposed by Foldy \cite{foldy}.  For suitable choices of
the  coupling  constants  $\epsilon  _n$  and $\mu  _n$,  e.g.,  the  minimal
($\epsilon_0=e;  \ \epsilon_n=0, \ n\not=0;  \  \mu_n=0  \ \forall n$) or the
Pauli ($\epsilon_n=0 \ \forall n;  \ \mu_0=\chi\frac{e}{2m_0};   \ \mu_n=0, \
n\not=0$) couplings are obtained.  The Dirac equation that results, retaining
the  three  first  terms, has been used frequently in the past  in  order  to
describe nucleons.    Furthermore, we have proved recently that this equation
provides a model to obtain the relativistic part of the radiative corrections
of the energy levels  of  the  electron  in an external electromagnetic field
\cite{barker}.

Alternatively, if we choose

\begin{equation}
R=g \gamma ^\mu A_\mu^a L_a, \label{r1}
\end{equation} 
where $g$  is a coupling constant, $A_\mu^a$ are Yang-Mills fields, and $L_a$
the corresponding generators  of  the gauge group, $[L_a,L_b]=iC_{ab}^c L_c$,
we can reproduce the  equations  of motion of a ``colored'' spinning particle
\cite{heinz}, as we will show in another paper \cite{color}.  

Another possibility is \cite{bazeia}

\begin{equation}  
R=e\gamma ^\mu A_\mu +\mu_0^{(m)}\frac{e}{2m_0} \sigma^{\mu\nu}  F_{\mu\nu} +
\mu_0^{(e)} \frac{e}{2m_0} \gamma_5 \sigma^{\mu\nu} F_{\mu\nu}, \label{r2}
\end{equation}
which  takes into account particles with  anomalous  moments  \cite{feinberg}
(including a possible time reversal violation).  The last term can be thought
of as a Pauli coupling containing the dual  electromagnetic  tensor  (see the
first identity of footnote 6), i.e., the case of a monopole Pauli coupling.


\begin{references}
\bibitem{heinz}Some  applications  for strong interactions were treated by U.
Heinz,  Phys.    Lett.  {\bf 144B}, 228 (1984).   Our  results  can  also  be
extrapolated to this case.
\end{references}
\end{document}